\documentclass[english]{article}
\usepackage[T1]{fontenc}
\usepackage[latin9]{inputenc}
\usepackage{booktabs}
\usepackage{amsmath}
\usepackage{graphicx}

\makeatletter

\providecommand{\tabularnewline}{\\}

\newcommand{\lyxaddress}[1]{
\par {\raggedright #1
\vspace{1.4em}
\noindent\par}
}

\usepackage{listings}
\usepackage{microtype}
\usepackage[T1]{fontenc}

\makeatother

\usepackage{babel}
\usepackage{listings}

\begin{document}

\title{Direct Calculation of the Transfer Map of Electrostatic Deflectors,
and Comparison with the Codes \emph{COSY INFINITY} and \emph{GIOS}\\
\ \\
MSU Report MSUHEP 171023}

\author{Eremey Valetov, Martin Berz, and Kyoko Makino}

\date{November 1, 2017}
\maketitle
\begin{abstract}
\emph{COSY INFINITY} uses a beamline coordinate system with a Frenet-Serret
frame relative to the reference particle, and calculates differential
algebra\textendash valued transfer maps by integrating the ODEs of
motion in the respective vector space over a differential algebra
(DA).

We will describe and perform computation of the DA transfer map of
an electrostatic spherical deflector in a laboratory coordinate system
using two conventional methods: (1) by integrating the ODEs of motion
using a 4th order Runge-Kutta integrator and (2) by computing analytically
and in closed form the properties of the respective elliptical orbits
from Kepler theory. We will compare the resulting transfer maps with
(3) the DA transfer map of \emph{COSY INFINITY}'s built-in electrostatic
spherical deflector element $\mathtt{ESP}$ and (4) the transfer map
of the electrostatic spherical deflector computed using the program
\emph{GIOS}.

In addition to the electrostatic spherical deflector, we study an
electrostatic cylindrical deflector, where the Kepler theory is not
applicable. We compute the DA transfer map by the ODE integration
method (1), and compare it with the transfer maps by (3) \emph{COSY
INFINITY}'s built-in electrostatic cylindrical deflector element $\mathtt{ECL}$,
and (4) \emph{GIOS}.

In addition to the code listings in the appendices, the codes to run
the test cases are available at

\noindent \texttt{http://bt.pa.msu.edu/cgi-bin/display.pl?name=ELSPHTM17}
\end{abstract}

\lyxaddress{Michigan State University, East Lansing, MI 48824, USA}

\section{Introduction}

\emph{COSY INFINITY} \cite{COSY} uses a beamline coordinate system
that comprises the phase space coordinates \cite[p. 9]{COSY-BPM-10}
\begin{center}
\begin{tabular}{ll}
$r_{1}=x,$ & $r_{2}=a=p_{x}/p_{0},$\tabularnewline
$r_{3}=y,$ & $r_{4}=b=p_{y}/p_{0},$\tabularnewline
$r_{5}=l=-\left(t-t_{0}\right)v_{0}\frac{\gamma}{1+\gamma},$ & $r_{6}=\delta_{K}=\frac{K-K_{0}}{K}.$\tabularnewline
\end{tabular}
\par\end{center}

Coordinates $x$ and $y$ are the transversal Frenet-Serret position
coordinates in meters, $p$ is the momentum, $K$ is the kinetic energy,
$v$ is the velocity, $t$ is the time of flight, and $\gamma$ is
the Lorentz factor. The index $0$ refers to the reference particle.

In \emph{COSY INFINITY}, the equations of motion are integrated, once
for each particle optical element comprising the lattice, in phase
space represented as a vector space $\left(_{n}D_{v}\right)^{v}$
over a differential algebra (DA) $_{n}D_{v}$ \cite[pp. 86--100]{Berz1999},
where $n$ is the computation order and $v$ is the number of phase
space coordinates. The result is a transfer map $\mathcal{M}$ that
expresses the final coordinates $z_{\mathrm{f}}$ of any ray as $z_{\mathrm{f}}=\mathcal{M}\left(z_{\mathrm{i}}\right)$,
where $z_{\mathrm{i}}$ are the initial coordinates \cite[Chs. 4--5]{Berz1999}.

\emph{COSY INFINITY} includes a built-in electrostatic spherical deflector
particle optical element $\mathtt{ESP}$. We will describe two conventional
methods of computing the non-relativistic DA transfer map of an electrostatic
spherical deflector: (1) by integrating the ODEs of motion using a
4th order Runge-Kutta integrator and (2) by computing and applying
the transition matrix with elements as the Lagrange coefficients.
We will calculate the transfer map of an electrostatic spherical deflector
using these two methods and compare the results with (3) the DA transfer
map of the non-relativistic version of \emph{COSY INFINITY}'s built-in
electrostatic spherical deflector element $\mathtt{ESP}$ and (4)
the transfer map of the electrostatic spherical deflector computed
using the program \emph{GIOS} \cite{GIOS}.

Additionally, we perform a similar study on an electrostatic cylindrical
deflector, where the Kepler theory is not applicable, so we compute
and compare the transfer maps by the methods (1), (3) using $\mathtt{ECL}$,
and (4).

\section{Transfer Map Calculation by Integrating the ODEs in Laboratory Coordinates}

Consider a bunch of non-relativistic charged particles launched with
kinetic energy $K_{0}=mv_{0}^{2}/2$ and zero potential energy, where
$m$ is the particle mass and $v_{0}$ is the reference velocity.
Suppose that a circular reference orbit of radius $r_{0}$ is defined
for the particle bunch in an electrostatic deflector. Now, for concreteness,
consider one particle in this bunch. Following the convention of the
potential energy as zero at the reference orbit, we calibrate the
potential energy of the charged particle such that 
\[
U(r_{0})=0.
\]
In the following section, we discuss the specific case of an electrostatic
spherical deflector. The case of an electrostatic cylindrical deflector
in Sub-Sec. \ref{sec:ECL}.

\subsection{The ODEs in an Electrostatic Spherical Deflector\label{sec:ESP}}

The potential energy in an electrostatic spherical deflector is 
\begin{equation}
U\left(r\right)=-\frac{\alpha}{r}+\frac{\alpha}{r_{0}},\label{eq:U}
\end{equation}
where the electrostatic potential $V$ and the electric field are
\begin{equation}
U(r)=ZeV(r),\quad E_{r}(r)=-\frac{\partial V}{\partial r}=-\frac{\alpha}{Ze}\frac{1}{r^{2}}.\label{eq:Er}
\end{equation}
By energy conservation, an off-reference particle at initial radius
$r_{\mathrm{i}}$ would upon entering the deflector have an initial
velocity magnitude $v_{\mathrm{i}}$ such that
\begin{equation}
K_{0}=\frac{mv_{0}^{2}}{2}=\frac{mv_{\mathrm{i}}^{2}}{2}+U\left(r\right)=\frac{mv_{\mathrm{i}}^{2}}{2}-\frac{\alpha}{r_{\mathrm{i}}}+\frac{\alpha}{r_{0}},\label{eq:Econs}
\end{equation}
from where we obtain
\begin{equation}
v_{\mathrm{i}}=\sqrt{v_{0}^{2}-\frac{2\alpha}{m}\left(\frac{1}{r_{0}}-\frac{1}{r_{\mathrm{i}}}\right)},\label{eq:v1-v0}
\end{equation}
or, expressing $\alpha$ in terms of the particle charge $q=Ze$,
the reference orbit radius $r_{0}$, and the electric rigidity $\chi_{\mathrm{e}}=\frac{pv}{Ze}$
as $\alpha=Ze\chi_{\mathrm{e}}r_{0}$, 
\begin{equation}
v_{\mathrm{i}}=\sqrt{v_{0}^{2}-\frac{2}{m}Ze\chi_{\mathrm{e}}\left(1-\frac{r_{0}}{r_{\mathrm{i}}}\right)}.\label{eq:v1-velocity}
\end{equation}

\emph{COSY INFINITY}'s horizontal transversal coordinate $x$ is defined
relative to the circular reference orbit as $x=r-r_{0}$, where $r$
is the length of the projection of a particle's radius vector on the
plane of the reference particle's orbital motion. \emph{COSY INFINITY}'s
horizontal momentum component $a=p_{x}/p_{0}$ can be, in the non-relativistic
case, expressed as $a=v_{x}/v_{0}$, where $v_{x}$ is the $x$ component
of a particle's velocity and $v_{0}$ is the reference velocity.

Now, let $\left(x_{\mathrm{i}},a_{\mathrm{i}}\right)$ be the initial
beamline coordinates of the particle in the electrostatic spherical
deflector. In reality, the deflector would have a fringe field; in
this model, we approximate the fringe field by an instant jump in
the electrostatic potential from zero to the electrostatic potential
of the electrostatic spherical deflector at radius $r_{\mathrm{i}}=r_{0}+x_{\mathrm{i}}$.
Thus, all particles in a bunch experience a ``step down'' from kinetic
energy $K_{0}$ to a kinetic energy $K=K_{0}-U\left(r\right)$ at
time $t=0$.

In a similar way, at the end of the element, the particle will experience
a \textquotedbl{}step up\textquotedbl{} due to the change of the potential
energy back to zero. Note that these changes of energy are essential
even in the absence of a true fringe field treatment to preserve the
actual beam energy.

For calculations, we will use the polar laboratory coordinate system
$\left(r,\theta\right)$, with particles launched at polar angle $\theta=0$.
Additionally, we will use the Cartesian laboratory coordinate system
\[
\left(\tilde{x},\tilde{y}\right)=r\left(\cos\theta,\sin\theta\right).
\]

In this Cartesian laboratory coordinate system, the particle has the
initial position
\begin{equation}
\vec{r}_{\mathrm{i}}=\left(r_{0}+x_{\mathrm{i}},0\right)\label{eq:init-r}
\end{equation}
and the initial velocity
\begin{equation}
\vec{v}_{\mathrm{i}}=\left(a_{\mathrm{i}}v_{0},\sqrt{v_{\mathrm{i}}^{2}-\left(a_{\mathrm{i}}v_{0}\right)^{2}}\right),\label{eq:init-v}
\end{equation}
where $v_{\mathrm{i}}$ is the initial particle velocity magnitude
at the initial radius $r_{\mathrm{i}}$ obtained using eq. \ref{eq:v1-velocity}.

From the second Newton law, the radial acceleration of the particle
in the electrostatic spherical deflector is
\begin{equation}
\frac{d^{2}}{dt^{2}}r=-\frac{\mu}{r^{2}}+\omega^{2}r,\label{eq:ODE}
\end{equation}
where $\mu=\alpha/m$ and $\omega$ is the particle's angular velocity.

Considering eqns. \ref{eq:init-r} and \ref{eq:init-v}, the conservation
of the angular momentum can be expressed in terms of massless angular
momentum 
\[
h=\omega r^{2}=\left|\vec{r}\times\vec{v}\right|
\]
 as
\begin{equation}
\begin{alignedat}{1}h & =\left|\vec{r}_{\mathrm{i}}\times\vec{v}_{\mathrm{i}}\right|=\left|\left(\vec{r}_{\mathrm{i}}\right)_{1}\left(\vec{v}_{\mathrm{i}}\right)_{2}-\left(\vec{r}_{\mathrm{i}}\right)_{2}\left(\vec{v}_{\mathrm{i}}\right)_{1}\right|=\\
 & =\left|\left(r_{0}+x_{\mathrm{i}}\right)\sqrt{v_{\mathrm{i}}^{2}-\left(a_{\mathrm{i}}v_{0}\right)^{2}}-0\cdot a_{\mathrm{i}}v_{0}\right|=\\
 & =\left(r_{0}+x_{\mathrm{i}}\right)\sqrt{v_{\mathrm{i}}^{2}-\left(a_{\mathrm{i}}v_{0}\right)^{2}}.
\end{alignedat}
\label{eq:h-init-coord}
\end{equation}
Taking the time derivative of $\omega=h/r^{2}$, we obtain
\[
\frac{d^{2}}{dt^{2}}\theta=-2\frac{h}{r^{3}}v_{r}.
\]

The final position $\vec{r}_{\mathrm{f}}$ and velocity $\vec{v}_{\mathrm{f}}$
can be obtained by integrating the ODEs of motion in the polar coordinates
\begin{equation}
\frac{d}{dt}\left(\begin{array}{c}
r\\
v_{r}\\
\theta\\
\omega
\end{array}\right)=\left(\begin{array}{c}
v_{r}\\
-\frac{\mu}{r^{2}}+\omega^{2}r\\
\omega\\
-2\frac{h}{r^{3}}v_{r}
\end{array}\right),\label{eq:Kepler-ODEs-1}
\end{equation}
with the initial condition 
\[
\left(\begin{array}{c}
r\\
v_{r}\\
\theta\\
\omega
\end{array}\right)_{0}=\left(\begin{array}{c}
\left(\vec{r}_{\mathrm{i}}\right)_{1}\\
\left(\vec{v}_{\mathrm{i}}\right)_{1}\\
0\\
\left(\vec{v}_{\mathrm{i}}\right)_{2}/\left(\vec{r}_{\mathrm{i}}\right)_{1}
\end{array}\right).
\]

Considering that $d\theta/dt=\omega$, applying the chain rule, this
system of ODEs can be expressed in terms of polar angle $\theta$
as the independent variable as
\begin{equation}
\frac{d}{d\theta}\left(\begin{array}{c}
r\\
v_{r}\\
\theta\\
\omega
\end{array}\right)=\left(\begin{array}{c}
\frac{v_{r}}{\omega}\\
-\frac{\mu}{r^{2}}\frac{1}{\omega}+\omega r\\
1\\
-2\frac{h}{r^{3}}\frac{v_{r}}{\omega}
\end{array}\right)\label{eq:Kepler-ODEs-2}
\end{equation}
with the same initial condition.

Having obtained the final position 
\[
\vec{r}_{\mathrm{f}}=r_{\mathrm{f}}\left(\cos\theta_{\mathrm{f}},\sin\theta_{\mathrm{f}}\right)
\]
and velocity $\vec{v}_{\mathrm{f}}$, the final $\left(x,a\right)$
coordinates are
\[
\left(x_{\mathrm{f}},a_{\mathrm{f}}\right)=\left(r_{\mathrm{f}}-r_{0},\frac{\vec{v}_{\mathrm{f}}\cdot\vec{r}_{\mathrm{f}}}{v_{0}r_{\mathrm{f}}}\right).
\]

We calculated the DA transfer map of the electrostatic spherical deflector,
in $\left(x,a\right)$ beamline coordinates, with reference orbit
radius $r_{0}=1\:\mathrm{m}$, by integration of the ODEs of motion
in polar laboratory coordinates using a 4th order Runge-Kutta integrator.
This transfer map is listed in Sec. \ref{sec:Kepler-tm-comparison}.

To obtain a DA-valued transfer map, we set the initial phase space
coordinates as 
\[
\left(x_{\mathrm{i}},a_{\mathrm{i}}\right)=\left(d_{1},d_{2}\right)=\left(\mathtt{DA\left(1\right)},\mathtt{DA\left(2\right)}\right),
\]
the respective DA generators \cite[pp. 86--96]{Berz1999}. We performed
the calculations with the computation order $3$.

A version of the \emph{COSY INFINITY} program that uses a 4th order
Runge-Kutta integrator to obtain the transfer map by solving the ODEs
of motion in polar laboratory coordinates is listed in App. \ref{sec:code-ODE-COSY}.
Additionally, App. \ref{sec:code-ODE-Mathematica} lists a \emph{Mathematica}
program that integrates the ODEs in polar laboratory coordinates for
an individual orbit, plots one turn of the orbit, and outputs the
final $\left(x,a\right)$ coordinates.

\subsection{The ODEs in an Electrostatic Cylindrical Deflector\label{sec:ECL}}

The motion in an electrostatic cylindrical deflector is described
in a similar way as in the case of a spherical deflector. Corresponding
to eqns. \ref{eq:U} and \ref{eq:Er}, the potential energy in an
electrostatic cylindrical deflector is 
\begin{equation}
U\left(r\right)=ZeV(r)=\alpha\log\frac{r}{r_{0}},\label{eq:UCL}
\end{equation}
and the electric field is
\begin{equation}
E_{r}(r)=-\frac{\partial V}{\partial r}=-\frac{\alpha}{Ze}\frac{1}{r}.\label{eq:ErCL}
\end{equation}
The energy conservation equation \ref{eq:Econs} is replaced by
\begin{equation}
K_{0}=\frac{mv_{0}^{2}}{2}=\frac{mv_{\mathrm{i}}^{2}}{2}+U\left(r\right)=\frac{mv_{\mathrm{i}}^{2}}{2}+\alpha\log\frac{r_{i}}{r_{0}},\label{eq:EconsCL}
\end{equation}
resulting in the following initial velocity $v_{\mathrm{i}},$ instead
of the corresponding equations in \ref{eq:v1-v0} and \ref{eq:v1-velocity}.
\begin{equation}
v_{\mathrm{i}}=\sqrt{v_{0}^{2}-\frac{2\alpha}{m}\log\frac{r_{i}}{r_{0}}}=\sqrt{v_{0}^{2}-\frac{2}{m}Ze\chi_{\mathrm{e}}r_{0}\log\frac{r_{i}}{r_{0}}}.\label{eq:v1-velocityCL}
\end{equation}

Corresponding to the radial equation of motion in \ref{eq:ODE}, we
have in the electrostatic cylindrical deflector
\begin{equation}
\frac{d^{2}}{dt^{2}}r=-\frac{\mu}{r}+\omega^{2}r,\label{eq:ODECL}
\end{equation}
resulting in, instead of eq. \ref{eq:Kepler-ODEs-1},

\begin{equation}
\frac{d}{dt}\left(\begin{array}{c}
r\\
v_{r}\\
\theta\\
\omega
\end{array}\right)=\left(\begin{array}{c}
v_{r}\\
-\frac{\mu}{r}+\omega^{2}r\\
\omega\\
-2\frac{h}{r^{3}}v_{r}
\end{array}\right),\label{eq:ODEs-1CL}
\end{equation}
and arriving at a set of ODEs in terms of $\theta,$ corresponding
to eq. \ref{eq:Kepler-ODEs-2}, as
\begin{equation}
\frac{d}{d\theta}\left(\begin{array}{c}
r\\
v_{r}\\
\theta\\
\omega
\end{array}\right)=\left(\begin{array}{c}
\frac{v_{r}}{\omega}\\
-\frac{\mu}{r}\frac{1}{\omega}+\omega r\\
1\\
-2\frac{h}{r^{3}}\frac{v_{r}}{\omega}
\end{array}\right)\label{eq:ODEs-2CL}
\end{equation}
with the initial condition 
\[
\left(\begin{array}{c}
r\\
v_{r}\\
\theta\\
\omega
\end{array}\right)_{0}=\left(\begin{array}{c}
\left(\vec{r}_{\mathrm{i}}\right)_{1}\\
\left(\vec{v}_{\mathrm{i}}\right)_{1}\\
0\\
\left(\vec{v}_{\mathrm{i}}\right)_{2}/\left(\vec{r}_{\mathrm{i}}\right)_{1}
\end{array}\right),
\]
where $v_{i}$ expressed in eq. \ref{eq:v1-velocityCL} is to be used.
The other equations stay the same as in the electrostatic spherical
deflector case.

In a similar way to the electrostatic spherical deflector case, a
DA transfer map is calculated and is listed in Sec. \ref{sec:tmap-ECL}.
A \emph{COSY INFINITY} program is listed in App. \ref{sec:code-ODE-COSY}.

\section{Transfer Map Calculation Using Lagrange Coefficients in Laboratory
Coordinates\label{sec:Kepler}}

Motion in a central field with a potential energy of the form $U\left(r\right)=-\alpha/r+\mathrm{const}$
is described by conventional Kepler theory; in particular, by the
equation of orbit
\begin{equation}
r=\frac{p}{1+e\cos f},\label{eq:eqn-orbit}
\end{equation}
where $p$ is the focal parameter, $e$ is the orbit eccentricity,
and polar angle $f$ is called the true anomaly \cite[p. 117]{battin}.
True anomaly $f=0$ corresponds to the direction of the perihelion,
that is, the point of the orbit nearest to the orbital focus at the
origin of the polar coordinate system $\left(r,f\right)$.

\begin{figure}[t]
\begin{centering}
\includegraphics[width=0.8\textwidth]{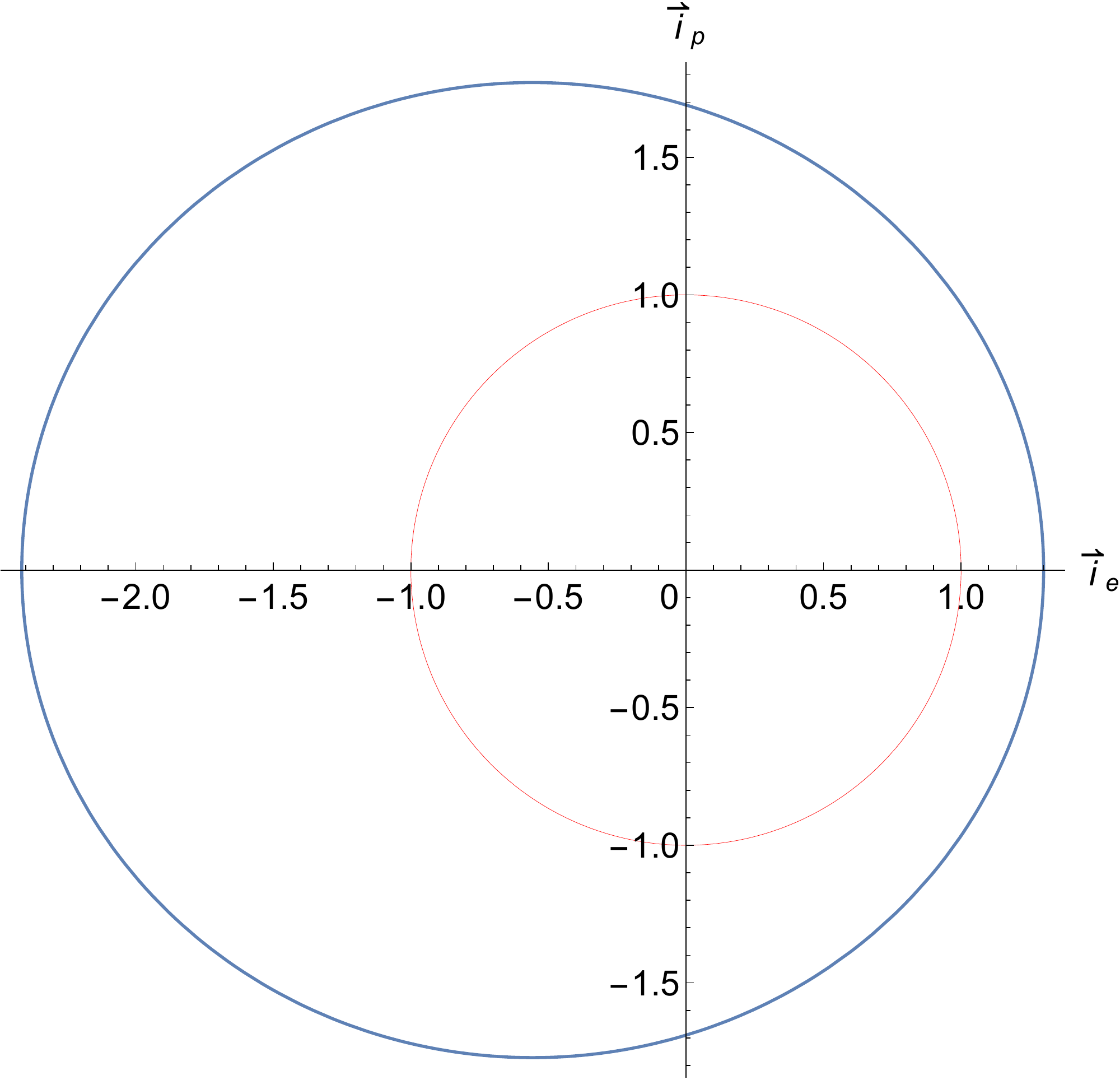}
\par\end{centering}
\caption{Orbit of a particle launched counter-clockwise at polar angle $f_{\mathrm{i}}=0$
with initial beamline coordinates $\left(x_{\mathrm{i}},a_{\mathrm{i}}\right)=\left(0.3,0\right)$
through an electrostatic spherical deflector (blue). The reference
orbit (red) has the radius $r_{0}=1\:\mathrm{m}$. The plot illustrates
the basis vectors $\left(\vec{i}_{e},\vec{i}_{p}\right)$ of the heliocentric
coordinate system in relation to the orbit geometry. The plot was
generated by the \emph{Mathematica} notebook for integration of the
ODEs for individual orbits in the polar laboratory coordinate system
listed in App. \ref{sec:code-ODE-Mathematica}.\label{fig:Kepler-coord}}
\end{figure}

The particle position $\vec{r}$ is expressed as
\begin{equation}
\vec{r}=\vec{i}_{e}r\cos f+\vec{i}_{p}r\sin f,\label{eq:r-orb}
\end{equation}
in terms of polar coordinates $\left(r,f\right)$ and the basis vectors
$\left(\vec{i}_{e},\vec{i}_{p}\right)$ of the heliocentric coordinate
system, where $\vec{i}_{e}$ is a unit vector in the direction of
perihelion and unit vector $\vec{i}_{p}$ is chosen so that $\vec{i}_{e}\times\vec{i}_{p}$
is co-directional with the vector of the angular velocity of the particle,
as Fig. \ref{fig:Kepler-coord} illustrates.

According to the second Newton law, zero radial acceleration at the
reference orbit requires
\[
\frac{mv_{0}^{2}}{r_{0}}=\frac{\alpha}{r_{0}^{2}}.
\]
Thus, the parameter $\mu=\alpha/m$ can be expressed as
\begin{equation}
\mu=\frac{\alpha}{m}=v_{0}^{2}r_{0}\label{eq:mu}
\end{equation}
in terms of the reference velocity $v_{0}$ and the reference orbit
radius $r_{0}$.

The focal parameter $p$ can be expressed in terms of $\mu$ and the
massless angular momentum $h=\left|\vec{r}\times\vec{v}\right|$ as
\begin{equation}
p=\frac{h^{2}}{\mu},\label{eq:focal-parameter}
\end{equation}
which is obtained in course of a standard derivation of the equation
of orbit, eq. \ref{eq:eqn-orbit}, as in \cite[pp. 114--116]{battin}.

We note that taking the time derivative of both sides of eq. \ref{eq:eqn-orbit}
gives
\begin{equation}
\dot{r}=\dot{f}r\frac{e\sin f}{1+e\cos f}.\label{eq:r-der}
\end{equation}

Considering eqns. \ref{eq:r-der} and \ref{eq:focal-parameter} and
conservation of massless angular momentum $h=\dot{f}r^{2}$, taking
the time derivative of the position vector $\vec{r}$ in eq. \ref{eq:r-orb}
yields for the particle velocity 
\begin{equation}
\begin{alignedat}{1}\vec{v} & =\vec{i}_{e}\left(\dot{r}\cos f-r\dot{f}\sin f\right)+\vec{i}_{p}\left(\dot{r}\sin f+r\dot{f}\cos f\right)=\\
 & =\vec{i}_{e}\frac{h}{r}\left(\frac{e\sin f}{1+e\cos f}\cos f-\sin f\right)+\vec{i}_{p}\frac{h}{r}\left(\frac{e\sin f}{1+e\cos f}\sin f+\cos f\right)=\\
 & =\vec{i}_{e}\frac{h}{r}\left(-\frac{\sin f}{1+e\cos f}\right)+\vec{i}_{p}\frac{h}{r}\left(\frac{e+\cos f}{1+e\cos f}\right)=\\
 & =-\vec{i}_{e}\frac{\mu}{h}\sin f+\vec{i}_{p}\frac{\mu}{h}\left(e+\cos f\right).
\end{alignedat}
\label{eq:v-orb}
\end{equation}

Rewriting eqns. \ref{eq:r-orb} and \ref{eq:v-orb} in matrix form,
we have
\begin{equation}
\left(\begin{array}{c}
\vec{r}\\
\vec{v}
\end{array}\right)=A\left(\begin{array}{c}
\vec{i}_{e}\\
\vec{i}_{p}
\end{array}\right),\quad\textrm{where }A=\left(\begin{array}{cc}
r\cos f & r\sin f\\
-\frac{\mu}{h}\sin f & \frac{\mu}{h}\left(e+\cos f\right)
\end{array}\right).\label{eq:r,v-vec}
\end{equation}
We note that the determinant of the matrix $A$ is, considering eq.
\ref{eq:focal-parameter},
\begin{equation}
\begin{alignedat}{1}\det A & =r\frac{\mu}{h}\left(e+\cos f\right)\cos f+r\frac{\mu}{h}\sin^{2}f=\\
 & =r\frac{\mu}{h}\left(1+e\cos f\right)=r\frac{\mu}{h}\frac{p}{r}=h.
\end{alignedat}
\label{eq:detA}
\end{equation}

Let $f_{\mathrm{i}}$ be the initial true anomaly of a particle in
the spherical electrostatic deflector. Then, from the equation of
orbit in eq. \ref{eq:eqn-orbit},
\begin{equation}
e\cos f_{\mathrm{i}}=\frac{p}{r_{\mathrm{i}}}-1,\label{eq:e-cos}
\end{equation}
where $r_{\mathrm{i}}$ is the initial radius.

Taking the scalar product of initial position $\vec{r}_{\mathrm{i}}$
and initial velocity $\vec{v}_{\mathrm{i}}$, we obtain from eq. \ref{eq:r,v-vec}
that
\begin{alignat*}{1}
\vec{r}_{\mathrm{i}}\cdot\vec{v}_{\mathrm{i}} & =-r_{\mathrm{i}}\frac{\mu}{h}\sin f_{\mathrm{i}}\cos f_{\mathrm{i}}+r_{\mathrm{i}}\frac{\mu}{h}\sin f_{\mathrm{i}}\left(e+\cos f_{\mathrm{i}}\right)=\\
 & =r_{\mathrm{i}}\frac{\mu}{h}e\sin f_{\mathrm{i}}.
\end{alignat*}
Hence,
\begin{equation}
e\sin f_{\mathrm{i}}=\frac{\sqrt{p}\sigma_{0}}{r_{\mathrm{i}}},\label{eq:e-sin}
\end{equation}
where $\sigma_{0}$ is defined as \cite[p. 130]{battin}
\[
\sigma_{0}=\frac{\vec{r}_{\mathrm{i}}\cdot\vec{v}_{\mathrm{i}}}{\sqrt{\mu}}.
\]

Let $\theta=f_{\mathrm{f}}-f_{\mathrm{i}}$ be the true anomaly difference
between the final and initial positions. Applying eqns. \ref{eq:e-cos}
and \ref{eq:e-sin} to the equation of orbit in eq. \ref{eq:eqn-orbit}
yields the final radius
\begin{equation}
\begin{alignedat}{1}r_{\mathrm{f}} & =\frac{p}{1+e\cos f_{\mathrm{f}}}=\frac{p}{1+e\cos\left(f_{\mathrm{i}}+\theta\right)}=\\
 & =\frac{p}{1+\cos\theta\cos f_{\mathrm{i}}-\sin\theta\sin f_{\mathrm{i}}}=\\
 & =\frac{p}{1+\left(\frac{p}{r_{\mathrm{i}}}-1\right)\cos\theta-\frac{\sqrt{p}\sigma_{0}}{r_{\mathrm{i}}}\sin\theta}=\\
 & =r_{\mathrm{i}}\frac{p}{r_{\mathrm{i}}+\left(p-r_{\mathrm{i}}\right)\cos\theta-\sqrt{p}\sigma_{0}\sin\theta}.
\end{alignedat}
\label{eq:rf-ri}
\end{equation}

Solving eq. \ref{eq:r,v-vec} for the basis vectors $\left(\vec{i}_{e},\vec{i}_{p}\right)$
of the heliocentric coordinate system in terms of the initial position
$\vec{r}_{\mathrm{i}}$, velocity $\vec{v}_{\mathrm{i}}$, and true
anomaly $f_{\mathrm{i}}$, considering that $\det A=h$ according
to eq. \ref{eq:detA}, we have
\begin{equation}
\begin{alignedat}{1}\left(\begin{array}{c}
\vec{i}_{e}\\
\vec{i}_{p}
\end{array}\right) & =A_{\mathrm{i}}^{-1}\left(\begin{array}{c}
\vec{r_{\mathrm{i}}}\\
\vec{v_{\mathrm{i}}}
\end{array}\right)=\frac{1}{\det A}\left(\begin{array}{cc}
\frac{\mu}{h}\left(e+\cos f_{\mathrm{i}}\right) & -r_{\mathrm{i}}\sin f_{\mathrm{i}}\\
\frac{\mu}{h}\sin f_{\mathrm{i}} & r_{\mathrm{i}}\cos f_{\mathrm{i}}
\end{array}\right)\left(\begin{array}{c}
\vec{r_{\mathrm{i}}}\\
\vec{v_{\mathrm{i}}}
\end{array}\right)=\\
 & =\left(\begin{array}{cc}
\frac{\mu}{h^{2}}\left(e+\cos f_{\mathrm{i}}\right) & -\frac{r_{\mathrm{i}}}{h}\sin f_{\mathrm{i}}\\
\frac{\mu}{h^{2}}\sin f_{\mathrm{i}} & \frac{r_{\mathrm{i}}}{h}\cos f_{\mathrm{i}}
\end{array}\right)\left(\begin{array}{c}
\vec{r_{\mathrm{i}}}\\
\vec{v_{\mathrm{i}}}
\end{array}\right).
\end{alignedat}
\label{eq:ie-ip}
\end{equation}

An analytic expression of the final position $\vec{r}_{\mathrm{f}}$
and velocity $\vec{v}_{\mathrm{f}}$ in terms of the initial position
$\vec{r}_{\mathrm{i}}$ and velocity $\vec{v}_{\mathrm{i}}$ is obtained
by inserting eq. \ref{eq:ie-ip} in eq. \ref{eq:r,v-vec}, which gives
\begin{equation}
\begin{alignedat}{1}\left(\begin{array}{c}
\vec{r}_{\mathrm{f}}\\
\vec{v}_{\mathrm{f}}
\end{array}\right) & =A_{\mathrm{f}}A_{\mathrm{i}}^{-1}\left(\begin{array}{c}
\vec{r_{\mathrm{i}}}\\
\vec{v_{\mathrm{i}}}
\end{array}\right)=\left(\begin{array}{cc}
r_{\mathrm{f}}\cos f_{\mathrm{f}} & r_{\mathrm{f}}\sin f_{\mathrm{f}}\\
-\frac{\mu}{h}\sin f_{\mathrm{f}} & \frac{\mu}{h}\left(e+\cos f_{\mathrm{f}}\right)
\end{array}\right)\cdot\\
 & \cdot\left(\begin{array}{cc}
\frac{\mu}{h^{2}}\left(e+\cos f_{\mathrm{i}}\right) & -\frac{r_{\mathrm{i}}}{h}\sin f_{\mathrm{i}}\\
\frac{\mu}{h^{2}}\sin f_{\mathrm{i}} & \frac{r_{\mathrm{i}}}{h}\cos f_{\mathrm{i}}
\end{array}\right)\left(\begin{array}{c}
\vec{r_{\mathrm{i}}}\\
\vec{v_{\mathrm{i}}}
\end{array}\right)=\\
 & =\left(\begin{array}{cc}
F & G\\
F_{t} & G_{t}
\end{array}\right)\left(\begin{array}{c}
\vec{r_{\mathrm{i}}}\\
\vec{v_{\mathrm{i}}}
\end{array}\right).
\end{alignedat}
\label{eq:pre-trans-mat}
\end{equation}
The matrix on the left is usually referred to as the transition matrix.
Note that this is not the same as the so-called transfer matrix of
beam physics, because the transition matrix contains all nonlinear
effects by virtue of its elements on orbit parameters. Specifically,
we obtain for the transition matrix elements: the transition matrix
element $F$ is 
\begin{alignat*}{1}
F & =\frac{\mu}{h^{2}}r_{\mathrm{f}}\left[\cos f_{\mathrm{f}}\left(e+\cos f_{\mathrm{i}}\right)+\sin f_{\mathrm{f}}\sin f_{\mathrm{i}}\right]=\\
 & =\frac{\mu}{h^{2}}r_{\mathrm{f}}\left(e\cos f_{\mathrm{f}}+\cos\theta\right)=\\
 & =\frac{\mu}{h^{2}}r_{\mathrm{f}}\left[\left(\frac{p}{r_{\mathrm{f}}}-1\right)+\cos\theta\right]=\\
 & =1-\frac{r_{\mathrm{f}}}{p}\left(1-\cos\theta\right),
\end{alignat*}
the transition matrix element $F_{t}$ is
\begin{alignat*}{1}
F_{t} & =\frac{\mu^{2}}{h^{3}}\left[-\sin f_{\mathrm{f}}\left(e+\cos f_{\mathrm{i}}\right)+\left(e+\cos f_{\mathrm{f}}\right)\sin f_{\mathrm{i}}\right]=\\
 & =\frac{\mu^{2}}{h^{3}}\left[e\left(\sin f_{\mathrm{i}}-\sin f_{\mathrm{f}}\right)-\sin\theta\right]=\\
 & =\frac{\mu^{2}}{h^{3}}\left[e\left(\sin f_{\mathrm{i}}-\sin f_{\mathrm{i}}\cos\theta-\sin\theta\cos f_{\mathrm{i}}\right)-\sin\theta\right]=\\
 & =\frac{\mu^{2}}{h^{3}}\left[e\sin f_{\mathrm{i}}\left(1-\cos\theta\right)-\sin\theta\left(1+e\cos f_{\mathrm{i}}\right)\right]=\\
 & =\frac{\mu^{2}}{h^{3}}\left[\frac{\sqrt{p}\sigma_{0}}{r_{\mathrm{i}}}\left(1-\cos\theta\right)-\frac{p}{r_{\mathrm{i}}}\sin\theta\right]=\\
 & =\frac{\sqrt{\mu}}{r_{\mathrm{i}}p}\left[\sigma_{0}\left(1-\cos\theta\right)-\sqrt{p}\sin\theta\right],
\end{alignat*}
the transition matrix element $G$ is
\begin{alignat*}{1}
G & =\frac{1}{h}r_{\mathrm{f}}r_{\mathrm{i}}\left(-\cos f_{\mathrm{f}}\sin f_{\mathrm{i}}+\sin f_{\mathrm{f}}\cos f_{\mathrm{i}}\right)=\\
 & =\frac{r_{\mathrm{f}}r_{\mathrm{i}}}{h}\sin\theta=\frac{r_{\mathrm{f}}r_{\mathrm{i}}}{\sqrt{\mu p}}\sin\theta,
\end{alignat*}
and the transition matrix element $G_{t}$ is
\begin{alignat*}{1}
G_{t} & =\frac{\mu}{h^{2}}r_{\mathrm{i}}\left[\sin f_{\mathrm{f}}\sin f_{\mathrm{i}}+\left(e+\cos f_{\mathrm{f}}\right)\cos f_{\mathrm{i}}\right]=\\
 & =\frac{\mu}{h^{2}}r_{\mathrm{i}}\left(e\cos f_{\mathrm{i}}+\cos\theta\right)=\\
 & =\frac{\mu}{h^{2}}r_{\mathrm{i}}\left[\left(\frac{p}{r_{\mathrm{i}}}-1\right)+\cos\theta\right]=\\
 & =1-\frac{r_{\mathrm{i}}}{p}\left(1-\cos\theta\right).
\end{alignat*}

Thus, we obtained a transition matrix 
\[
\Phi=\left(\begin{array}{cc}
F & G\\
F_{t} & G_{t}
\end{array}\right)
\]
that expresses the final coordinates $\left(\vec{r}_{\mathrm{f}},\vec{v}_{\mathrm{f}}\right)$
as a function of the initial coordinates $\left(\vec{r}_{\mathrm{i}},\vec{v}_{\mathrm{i}}\right)$
as
\begin{equation}
\left(\begin{array}{c}
\vec{r}_{\mathrm{f}}\\
\vec{v}_{\mathrm{f}}
\end{array}\right)=\Phi\left(\begin{array}{c}
\vec{r}_{\mathrm{i}}\\
\vec{v}_{\mathrm{i}}
\end{array}\right)\label{eq:tmap-lagrange}
\end{equation}
and comprises elements \cite[pp. 128--131]{battin}
\begin{alignat*}{1}
F\left(\theta\right) & =1-\frac{r_{\mathrm{f}}}{p}\left(1-\cos\theta\right),\\
F_{t}\left(\theta\right) & =\frac{\sqrt{\mu}}{r_{\mathrm{i}}p}\left[\sigma_{0}\left(1-\cos\theta\right)-\sqrt{p}\sin\theta\right],\\
G\left(\theta\right) & =\frac{r_{\mathrm{f}}r_{\mathrm{i}}}{\sqrt{\mu p}}\sin\theta,\\
G_{t}\left(\theta\right) & =1-\frac{r_{\mathrm{i}}}{p}\left(1-\cos\theta\right).
\end{alignat*}
The elements $F$, $F_{t}$, $G$, and $G_{t}$ of the transition
matrix $\Phi$ are called Lagrange coefficients. The Lagrange coefficients
$F_{t}$ and $G_{t}$ are simply time derivatives of $F$ and $G$,
respectively.

Applying the transfer matrix from eq. \ref{eq:tmap-lagrange} to the
initial position $\vec{v}_{\mathrm{i}}$ and velocity $\vec{p}_{\mathrm{i}}$
from eqns. \ref{eq:init-r} and \ref{eq:init-v}, we obtain the final
position $\vec{r}_{\mathrm{f}}$ and velocity $\vec{v}_{\mathrm{f}}$.

The final $\left(x,a\right)$ coordinates are obtained from the final
position 
\[
\vec{r}_{\mathrm{f}}=r_{\mathrm{f}}\left(\cos\theta_{\mathrm{f}},\sin\theta_{\mathrm{f}}\right)
\]
 and velocity $\vec{v}_{\mathrm{f}}$ as
\[
\left(x_{\mathrm{f}},a_{\mathrm{f}}\right)=\left(r_{\mathrm{f}}-r_{0},\frac{\vec{v}_{\mathrm{f}}\cdot\vec{r}_{\mathrm{f}}}{v_{0}r_{\mathrm{f}}}\right).
\]

We calculated the DA transfer map of the electrostatic spherical deflector,
in $\left(x,a\right)$ beamline coordinates, with reference orbit
radius $r_{0}=1\:\mathrm{m}$, using the transition matrix with elements
as the Lagrange coefficients in terms of the true anomaly difference.
This transfer map is listed in Sec. \ref{sec:Kepler-tm-comparison}.

The transfer map only depends on reference orbit radius $r_{0}$ and
the central angle of the tracked sector of the electrostatic spherical
deflector. As long as the reference orbit radius $r_{0}$ is kept
the same by adjusting the charge of the inner sphere of the deflector,
the transfer map does not depend on the charged particle's kinetic
energy, mass, or charge. Indeed, considering eqns. \ref{eq:focal-parameter},
\ref{eq:h-init-coord}, \ref{eq:init-r}, and \ref{eq:mu}, the focal
parameter $p$ is
\begin{alignat*}{1}
p & =\frac{h^{2}}{\mu}=\frac{\left|\vec{r}_{\mathrm{i}}\times\vec{v}_{\mathrm{i}}\right|^{2}}{v_{0}^{2}r_{0}}=\\
 & =\frac{\left(r_{0}+x_{\mathrm{i}}\right)^{2}\left[v_{\mathrm{i}}^{2}-\left(a_{\mathrm{i}}v_{0}\right)^{2}\right]}{v_{0}^{2}r_{0}}=\\
 & =\frac{\left(r_{0}+x_{\mathrm{i}}\right)^{2}\left[v_{0}^{2}-\frac{2\alpha}{m}\left(\frac{1}{r_{0}}-\frac{1}{r_{0}+x_{\mathrm{i}}}\right)-\left(a_{\mathrm{i}}v_{0}\right)^{2}\right]}{v_{0}^{2}r_{0}}=\\
 & =\frac{\left(r_{0}+x_{\mathrm{i}}\right)^{2}\left[v_{0}^{2}-2v_{0}^{2}r_{0}\left(\frac{1}{r_{0}}-\frac{1}{r_{0}+x_{\mathrm{i}}}\right)-\left(a_{\mathrm{i}}v_{0}\right)^{2}\right]}{v_{0}^{2}r_{0}}=\\
 & =\frac{\left(r_{0}+x_{\mathrm{i}}\right)\left[2r_{0}-\left(1+a_{\mathrm{i}}^{2}\right)\left(r_{0}+x_{\mathrm{i}}\right)\right]}{r_{0}},
\end{alignat*}
and the eccentricity $e$ is \cite[p. 116]{battin} such that, considering
eqns. \ref{eq:init-r}, \ref{eq:v1-v0}, and \ref{eq:mu},
\begin{alignat*}{1}
1-e^{2} & =p\left(\frac{2}{r}-\frac{v^{2}}{\mu}\right)=p\left(\frac{2}{r_{0}+x_{\mathrm{i}}}-\frac{v_{\mathrm{i}}^{2}}{\mu}\right)=\\
 & =p\left[\frac{2}{r_{0}+x_{\mathrm{i}}}-\frac{v_{0}^{2}-2v_{0}^{2}r_{0}\left(\frac{1}{r_{0}}-\frac{1}{r_{0}+x_{\mathrm{i}}}\right)}{v_{0}^{2}r_{0}}\right]=\\
 & =p\left[\frac{2}{r_{0}+x_{\mathrm{i}}}-\frac{1-2r_{0}\left(\frac{1}{r_{0}}-\frac{1}{r_{0}+x_{\mathrm{i}}}\right)}{r_{0}}\right]=\\
 & =p\left[\frac{2}{r_{0}+x_{\mathrm{i}}}-\frac{2r_{0}-\left(r_{0}+x_{\mathrm{i}}\right)}{r_{0}\left(r_{0}+x_{\mathrm{i}}\right)}\right]=\frac{p}{r_{0}.}
\end{alignat*}
Thus, the focal parameter $p$ and the eccentricity $e$ depend only
on reference orbit radius $r_{0}$ and initial beamline coordinates
$\left(x_{\mathrm{i}},a_{\mathrm{i}}\right)$.

To obtain a DA-valued transfer map, we set the initial phase space
coordinates as 
\[
\left(x_{\mathrm{i}},a_{\mathrm{i}}\right)=\left(d_{1},d_{2}\right)=\left(\mathtt{DA\left(1\right)},\mathtt{DA\left(2\right)}\right),
\]
the respective DA generators \cite[pp. 86--96]{Berz1999}. We performed
the calculations with the computation order $3$. A \emph{COSY INFINITY}
program to obtain the transfer map is listed in App. \ref{sec:code-Kepler-COSY}.

\pagebreak{}

\section{Transfer Maps of an Electrostatic Spherical Deflector and Comparison\emph{\label{sec:Kepler-tm-comparison}}}

Here, we list and compare the DA transfer maps for particles passing
through a $45\textdegree$ sector of the electrostatic spherical deflector,
calculated
\begin{enumerate}
\item by integration of the ODEs of motion in polar laboratory coordinates
(eq. \ref{eq:Kepler-ODEs-2}) using a 4th order Runge-Kutta integrator;
\item using the transition matrix with elements as the Lagrange coefficients
in terms of the true anomaly difference;
\item for \emph{COSY INFINITY}'s built-in electrostatic spherical deflector
element $\mathtt{ESP}$; and
\item using the code sequence $\mathtt{E\:S}$ in the program \emph{GIOS}.
\end{enumerate}
In all cases, the reference orbit radius is $r_{0}=1\:\mathrm{m}$,
and non-relativistic equations of motion were used. For definiteness,
the particles in the bunch were set to kinetic energy $1\:\mathrm{MeV}$,
mass $1\:\mathrm{amu}$, and charge $1\:e$; however, as noted above,
this setting has no impact on the orbit geometry as long as the reference
orbit radius is kept the same by adjusting the voltages of the inner
and outer spherical shells of the electrostatic spherical deflector.
For visual transfer map comparison purposes, the computation order
$3$ was used in all cases, except for \emph{GIOS}, where the computation
order $2$ was used.

The Jacobian $M=\mathrm{Jac}\left(\mathcal{M}\right)$ of the transfer
map $\mathcal{M}$ of any Hamiltonian system satisfies the symplecticity
condition $M\cdot\hat{J}\cdot M^{\mathrm{T}}=\hat{J}$ \cite[pp. 155--159]{Berz1999},
where the phase space coordinates are ordered as $z=\left(q_{1},\ldots,q_{m},p_{1},\ldots,p_{m}\right),$
$\left(q_{1},\ldots,q_{m}\right)$ are the canonical position coordinates
and $\left(p_{1},\ldots,p_{m}\right)$ are the conjugate momentum
coordinates, 
\[
\hat{J}=\left(\begin{array}{cc}
0 & I_{m}\\
-I_{m} & 0
\end{array}\right),
\]
 $I_{m}$ is an $m\times m$ identity matrix, and $m$ is the phase
space dimension. 

For each transfer map calculation case, we computed deviations from
the conditions of symplecticity \cite[pp. 155--159]{Berz1999}\cite{wollnik-nimsys}
for the first and second order aberration coefficients and motion
in the $x\textrm{-}a$ plane. These conditions of symplecticity are
as follows:
\begin{equation}
\begin{array}{l}
g_{1}=\left(x|x\right)\left(a|a\right)-\left(a|x\right)\left(x|a\right)-1=0,\\
g_{2}=\left(x|x\right)\left(a|xa\right)-\left(a|x\right)\left(x|xa\right)+\left(x|xx\right)\left(a|a\right)-\left(a|xx\right)\left(x|a\right)=0,\\
g_{3}=\left(x|x\right)\left(a|aa\right)-\left(a|x\right)\left(x|aa\right)+\left(x|xa\right)\left(a|a\right)-\left(a|xa\right)\left(x|a\right)=0,
\end{array}\label{eq:sympl-cond}
\end{equation}
where the aberration coefficient $\left(z_{i}|z_{j_{1}}\cdots z_{j_{n}}\right)$
is the partial derivative
\[
\left(z_{i}|z_{j_{1}}\cdots z_{j_{n}}\right)=\left(\frac{\partial^{n}\left(\mathcal{M}\left(z\right)\right)_{i}}{\partial z_{j_{1}}\cdots\partial z_{j_{n}}}\right)_{z=0}
\]
of the $i$-th component of the respective transfer map $\mathcal{M}$
applied to a coordinate vector $z=\left(z_{1}\cdots z_{2m}\right)$,
and $2m$ is the number of phase space coordinates.\pagebreak{}

\subsection{Transfer Map Obtained by Integrating the ODEs in Laboratory Coordinates}

The transfer map from integration of the ODEs of motion in polar laboratory
coordinates using a 4th order Runge-Kutta integrator is as follows.
\begin{lstlisting}[tabsize=4,frame=shadowbox,mathescape=true]
TRANSFER MAP OBTAINED IN LAB COORDINATES
BY INTEGRATION OF THE ODEs OF MOTION
X_f
     I  COEFFICIENT            ORDER EXPONENTS
     1  0.7071067811865617       1   1 0  0
     2  0.7071067811865485       1   0 1  0
     3  -.4999999999998140       2   2 0  0
     4  0.9999999999999932       2   1 1  0
     5  0.2071067811865112       2   0 2  0
     6  -.3535533905929528       3   3 0  0
     7  0.6066017177973747E-01   3   1 2  0
     8  0.2928932188133920       3   0 3  0
     --------------------------------------
A_f
     I  COEFFICIENT            ORDER EXPONENTS
     1  -.7071067811864190       1   1 0  0
     2  0.7071067811865486       1   0 1  0
     3  -.7071067811865216       2   0 2  0
     4  -.3535533905931392       3   3 0  0
     5  -1.060660171779751       3   1 2  0
     --------------------------------------
\end{lstlisting}
DA coefficients with absolute values less than $10^{-11}$ are omitted. 

The deviations $g_{1}$, $g_{2}$, and $g_{3}$ from the conditions
of symplecticity listed in eq. \ref{eq:sympl-cond} in this case were
as follows:

\[
\begin{array}{l}
g_{1}=-0.7949196856316121\times10^{-13},\\
g_{2}=-0.2514743315828557\times10^{-12},\\
g_{3}=-0.1188442801551391\times10^{-12}.
\end{array}
\]

The 4th order Runge-Kutta integrator code for calculation of this
transfer map is listed in App. \ref{sec:code-ODE-COSY}.\pagebreak{}

\subsection{Transfer Map Obtained Using Lagrange Coefficients in Laboratory Coordinates}

The transfer map obtained in laboratory coordinates using the Lagrange
coefficients transition matrix is as follows.
\begin{lstlisting}[tabsize=4,frame=shadowbox,mathescape=true]
TRANSFER MAP OBTAINED IN LAB COORDINATES
USING LAGRANGE COEFFICIENTS
X_f
     I  COEFFICIENT            ORDER EXPONENTS
     1  0.7071067811865475       1   1 0  0
     2  0.7071067811865475       1   0 1  0
     3  -.5000000000000000       2   2 0  0
     4   1.000000000000000       2   1 1  0
     5  0.2071067811865475       2   0 2  0
     6  -.3535533905932737       3   3 0  0
     7  0.6066017177982122E-01   3   1 2  0
     8  0.2928932188134523       3   0 3  0
     --------------------------------------
A_f
     I  COEFFICIENT            ORDER EXPONENTS
     1  -.7071067811865475       1   1 0  0
     2  0.7071067811865476       1   0 1  0
     3  -.7071067811865475       2   0 2  0
     4  -.3535533905932737       3   3 0  0
     5  -1.060660171779821       3   1 2  0
     --------------------------------------
\end{lstlisting}
DA coefficients with absolute values less than $10^{-11}$ are omitted. 

The deviations $g_{1}$, $g_{2}$, and $g_{3}$ from the conditions
of symplecticity listed in eq. \ref{eq:sympl-cond} in this case were
as follows:

\[
\begin{array}{l}
g_{1}=-0.1110223024625157\times10^{-15},\\
g_{2}=0.3119771259853192\times10^{-16},\\
g_{3}=0.1110223024625157\times10^{-15}.
\end{array}
\]

A \emph{COSY INFINITY} code for calculation of this transfer map is
listed in App. \ref{sec:code-Kepler-COSY}.\pagebreak{}

\subsection{Transfer Map of \emph{COSY INFINITY}'s Built-In Electrostatic Spherical
Deflector Element $\mathtt{ESP}$}

The transfer map of \emph{COSY INFINITY}'s built-in electrostatic
spherical deflector element $\mathtt{ESP}$, obtained using non-relativistic
equations of motion, is as follows.
\begin{lstlisting}[tabsize=4,frame=shadowbox,mathescape=true]
TRANSFER MAP OF COSY INFINITY'S ESP ELEMENT

X_f
     I  COEFFICIENT            ORDER EXPONENTS
     1  0.7071067811865475       1   1 0  0
     2  0.7071067811865475       1   0 1  0
     3  -.4999999999999999       2   2 0  0
     4   1.000000000000000       2   1 1  0
     5  0.2071067811865475       2   0 2  0
     6  -.3535533905932738       3   3 0  0
     7  0.6066017177982123E-01   3   1 2  0
     8  0.2928932188134525       3   0 3  0
     --------------------------------------
A_f
     I  COEFFICIENT            ORDER EXPONENTS
     1  -.7071067811865475       1   1 0  0
     2  0.7071067811865475       1   0 1  0
     3  -.7071067811865475       2   0 2  0
     4  -.3535533905932737       3   3 0  0
     5  -1.060660171779821       3   1 2  0
     --------------------------------------
\end{lstlisting}

The deviations $g_{1}$, $g_{2}$, and $g_{3}$ from the conditions
of symplecticity listed in eq. \ref{eq:sympl-cond} were as follows:

\[
\begin{array}{l}
g_{1}=-0.2220446049250313\times10^{-15},\\
g_{2}=0.2220446049250313\times10^{-15},\\
g_{3}=0.3330669073875470\times10^{-15}.
\end{array}
\]

The \emph{COSY INFINITY} code for calculation of this transfer map
of the $\mathtt{ESP}$ element is listed in App. \ref{sec:code-ESP}.
We note that for simplicity, essentially the same result can be obtained
using the relativistic equations of motion that are by default used
in \textit{COSY INFINITY} by simply using a very low kinetic energy
for the calculation. For example, the transfer map computed using
the relativistic equations with the kinetic energy $10^{-7}$ agrees
with the above listed transfer map to about $10^{-10}.$ In this case,
the values of the deviations $g_{1}$, $g_{2}$, and $g_{3}$ are
$\sim10^{-15}.$\pagebreak{}

\subsection{Transfer Map Computed Using the Electrostatic Spherical Deflector
Element in the Program \emph{GIOS}}

The 2nd order transfer map of the electrostatic spherical deflector
computed using the code sequence $\mathtt{E\:S}$ in the program \emph{GIOS}
is as follows.
\begin{lstlisting}[tabsize=4,frame=shadowbox,mathescape=true]
TRANSFER MAP COMPUTED USING THE PROGRAM GIOS

X_f
     I  COEFFICIENT            ORDER EXPONENTS
     1  0.7071067812             1   1 0  0
     2  0.7071067812             1   0 1  0
     3  -.5000000000             2   2 0  0
     4  0.2928932188             2   1 1  0
     5  0.2071067812             2   0 2  0



     --------------------------------------
A_f
     I  COEFFICIENT            ORDER EXPONENTS
     1  -.7071067812             1   1 0  0
     2  0.7071067812             1   0 1  0
     3  -.5000000000             2   2 0  0
     4  -.7071067812             2   1 1  0
     5  -.2071067812             2   0 2  0
     --------------------------------------
\end{lstlisting}

The deviations $g_{1}$, $g_{2}$, and $g_{3}$ from the conditions
of symplecticity listed in eq. \ref{eq:sympl-cond} were as follows:

\[
\begin{array}{l}
g_{1}=0.3804934145534844\times10^{-10},\\
g_{2}=-0.2928932188380493,\\
g_{3}=0.7071067812000000.
\end{array}
\]
We note that the deviations $g_{2}$ and $g_{3}$ are significant
in magnitude and indicate error(s) in the program \emph{GIOS.}

The \emph{GIOS} input for calculation of this transfer map is listed
in App. \ref{sec:code-GIOS}. We also note that these differences
are not due to the fact that \emph{GIOS} uses momentum-like coordinates
that differ from those of \emph{COSY INFINITY}. The respective effects
manifest themselves only in order three in $x$ and $a$ terms, which
we are not comparing here. \emph{\pagebreak{}}

\section{Transfer Maps of an Electrostatic Cylindrical Deflector and Comparison\label{sec:tmap-ECL}}

In this section, we list and compare the DA transfer maps for particles
passing through a $45\textdegree$ sector of the electrostatic cylindrical
deflector, calculated
\begin{enumerate}
\item by integration of the ODEs of motion in polar laboratory coordinates
(eq. \ref{eq:ODEs-2CL}) using a 4th order Runge-Kutta integrator;
\item for \emph{COSY INFINITY}'s built-in electrostatic spherical deflector
element $\mathtt{ECL}$; and
\item using the code sequence $\mathtt{E\:S}$ in the program \emph{GIOS}.
\end{enumerate}
In all cases, the reference orbit radius is $r_{0}=1\:\mathrm{m}$,
and non-relativistic equations of motion were used. For definiteness,
the particles in the bunch were set to kinetic energy $1\:\mathrm{MeV}$,
mass $1\:\mathrm{amu}$, and charge $1\:e$; however, as noted above,
this setting has no impact on the orbit geometry as long as the reference
orbit radius is kept the same by adjusting the voltages of the inner
and outer cylindrical shells of the electrostatic cylindrical deflector.
For visual transfer map comparison purposes, the computation order
$3$ was used in all cases, except for \emph{GIOS}, where the computation
order $2$ was used.

\pagebreak{}

\subsection{Transfer Map Obtained by Integrating the ODEs in Laboratory Coordinates}

The transfer map from integration of the ODEs of motion in polar laboratory
coordinates using a 4th order Runge-Kutta integrator is as follows.
\begin{lstlisting}[tabsize=4,frame=shadowbox,mathescape=true]
TRANSFER MAP OBTAINED IN LAB COORDINATES
BY INTEGRATION OF THE ODEs OF MOTION
X_f
     I  COEFFICIENT            ORDER EXPONENTS
     1  0.4440158403262955       1   1 0  0
     2  0.6335810656653742       1   0 1  0
     3  -1.029322282408186       2   2 0  0
     4  0.4452197131126942       2   1 1  0
     5  0.9767302144878563E-01   2   0 2  0
     6  -.9310536195453081       3   3 0  0
     7  -.7814348139392595       3   2 1  0
     8  -.7214969045085385       3   1 2  0
     9  0.1172683765076191       3   0 3  0
     -------------------------------------- 
A_f
     I  COEFFICIENT            ORDER EXPONENTS
     1  -1.267162131330700       1   1 0  0
     2  0.4440158403261798       1   0 1  0
     3  -.3987403747459493       2   2 0  0
     4  -.3499052358015541       2   1 1  0
     5  -.7510014111250763       2   0 2  0
     6  -.6758776475462285       3   3 0  0
     7  -.2919765941780568       3   2 1  0
     8  -1.233526213798069       3   1 2  0
     9  -.2301781799921400       3   0 3  0
     --------------------------------------
\end{lstlisting}

The deviations $g_{1}$, $g_{2}$, and $g_{3}$ from the conditions
of symplecticity listed in eq. \ref{eq:sympl-cond} in this case were
as follows:

\[
\begin{array}{l}
g_{1}=-0.7316369732279782\times10^{-13},\\
g_{2}=0.1615374500829603\times10^{-12},\\
g_{3}=-0.2078615057854449\times10^{-12}.
\end{array}
\]

The 4th order Runge-Kutta integrator code for calculation of this
transfer map is listed in App. \ref{sec:code-ODE-COSY}.\pagebreak{}

\subsection{Transfer Map of \emph{COSY INFINITY}'s Built-In Electrostatic Cylindrical
Deflector Element $\mathtt{ECL}$}

The transfer map of \emph{COSY INFINITY}'s built-in electrostatic
spherical deflector element $\mathtt{ECL}$, obtained using non-relativistic
equations of motion, is as follows.
\begin{lstlisting}[tabsize=4,frame=shadowbox,mathescape=true]
TRANSFER MAP OF COSY INFINITY'S ESP ELEMENT

X_f
     I  COEFFICIENT            ORDER EXPONENTS
     1  0.4440158403262133       1   1 0  0
     2  0.6335810656653997       1   0 1  0
     3  -1.029322282408272       2   2 0  0
     4  0.4452197131126671       2   1 1  0
     5  0.9767302144879608E-01   2   0 2  0
     6  -.9310536195454117       3   3 0  0
     7  -.7814348139394898       3   2 1  0
     8  -.7214969045085790       3   1 2  0
     9  0.1172683765076182       3   0 3  0   
     --------------------------------------
A_f
     I  COEFFICIENT            ORDER EXPONENTS
     1  -1.267162131330799       1   1 0  0
     2  0.4440158403262133       1   0 1  0
     3  -.3987403747459333       2   2 0  0
     4  -.3499052358016756       2   1 1  0
     5  -.7510014111251326       2   0 2  0
     6  -.6758776475462280       3   3 0  0
     7  -.2919765941781459       3   2 1  0
     8  -1.233526213798173       3   1 2  0
     9  -.2301781799921575       3   0 3  0  
     --------------------------------------
\end{lstlisting}

The deviations $g_{1}$, $g_{2}$, and $g_{3}$ from the conditions
of symplecticity listed in eq. \ref{eq:sympl-cond} were as follows:

\[
\begin{array}{l}
g_{1}=0.4440892098500626\times10^{-15},\\
g_{2}=0.2220446049250313\times10^{-15},\\
g_{3}=0.2498001805406602\times10^{-15}.
\end{array}
\]

The \emph{COSY INFINITY} code for calculation of this transfer map
of the $\mathtt{ECL}$ element is listed in App. \ref{sec:code-ESP}.
We note that for simplicity, essentially the same result can be obtained
using the relativistic equations of motion that are by default used
in \textit{COSY INFINITY} by simply using a very low kinetic energy
for the calculation. For example, the transfer map computed using
the relativistic equations with the kinetic energy $10^{-7}$ agrees
with the above listed transfer map to about $10^{-10}.$ In this case,
the values of the deviations $g_{1}$, $g_{2}$, and $g_{3}$ are
$\sim10^{-15}.$\pagebreak{}

\subsection{Transfer Map Computed Using the Electrostatic Cylindrical Deflector
Element in the Program \emph{GIOS}}

The 2nd order transfer map of the electrostatic cylindrical deflector
computed using the code sequence $\mathtt{E\:S}$ in the program \emph{GIOS}
is as follows.
\begin{lstlisting}[tabsize=4,frame=shadowbox,mathescape=true]
TRANSFER MAP COMPUTED USING THE PROGRAM GIOS

X_f
     I  COEFFICIENT            ORDER EXPONENTS
     1  0.4440158574             1   1 0  0
     2  0.6335810705             1   0 1  0
     3  -1.029322250             2   2 0  0
     4  -.1883613764             2   1 1  0
     5  0.9767302679E-01         2   0 2  0




     --------------------------------------
A_f
     I  COEFFICIENT            ORDER EXPONENTS
     1  -1.267162098             1   1 0  0
     2  0.4440158574             1   0 1  0
     3  -.9613804652             2   2 0  0
     4  -1.399620967             2   1 1  0
     5  -.4696813690             2   0 2  0
     --------------------------------------
\end{lstlisting}

The deviations $g_{1}$, $g_{2}$, and $g_{3}$ from the conditions
of symplecticity listed in eq. \ref{eq:sympl-cond} were as follows:

\[
\begin{array}{l}
g_{1}=0.1705231511550664\times10^{-9},\\
g_{2}=-0.5559841747496004,\\
g_{3}=0.6335810760905867.
\end{array}
\]
We note that the deviations $g_{2}$ and $g_{3}$ are significant
in magnitude and indicate error(s) in the program \emph{GIOS. }

The \emph{GIOS} input for calculation of this transfer map is listed
in App. \ref{sec:code-GIOS}. We also note that these differences
are not due to the fact that \emph{GIOS} uses momentum-like coordinates
that differ from those of \emph{COSY INFINITY}. The respective effects
manifest themselves only in order three in $x$ and $a$ terms, which
we are not comparing here. \emph{\pagebreak{}}

\section{Calculation Results, Comparison, and Conclusion}

The electrostatic spherical deflector transfer maps calculated in
laboratory coordinates by integration of the ODEs and using the Lagrange
coefficients transition matrix are in very high agreement with the
transfer map of \emph{COSY INFINITY}'s built-in electrostatic spherical
deflector element $\mathtt{ESP}$. The deviations from the conditions
of symplecticity $g_{1}$, $g_{2}$, and $g_{3}$ were the highest
at $\sim10^{-13}$ in case of integration of the ODEs in polar laboratory
coordinates, the lowest at $\sim10^{-16}$ in case of calculation
using the Lagrange coefficients, and at $\sim10^{-16}$ in case of
the built-in element $\mathtt{ESP}$.

The transfer map of the electrostatic spherical deflector computed
using the program \emph{GIOS} significantly disagrees with the other
three transfer maps. However, the deviations $g_{2}$ and $g_{3}$
in the \emph{GIOS} case were also significant in magnitude and indicate
that the disagreement is due to error(s) in \emph{GIOS}.

We arrive at an equivalent conclusion for the electrostatic cylindrical
deflector transfer maps. The transfer map obtained by integrating
the ODEs in laboratory coordinates agrees well with those obtained
by \emph{COSY INFINITY}'s built-in electrostatic cylindrical deflector
element $\mathtt{ECL}$, both satisfying the symplectic conditions.
On the other hand, the transfer map obtained using \emph{GIOS} significantly
disagrees with the other two transfer maps, and also the symplectic
condition was significantly deviated.

\pagebreak{}

\appendix

\section{\emph{COSY INFINITY} Code for Transfer Map Calculation Using ODEs
in Laboratory Coordinates\label{sec:code-ODE-COSY}}

The following is a \emph{COSY INFINITY} code that calculates the transfer
map of a $45\textdegree$ sector of the electrostatic spherical deflector
in $\left(x,a\right)$ beamline coordinates by integrating the ODEs
of motion in polar laboratory coordinates. This version of the code
uses a 4th order Runge-Kutta integrator with fixed step size. The
inner sphere of the electrostatic spherical deflector is charged to
result in a circular reference orbit of radius $r_{0}=1\:\mathrm{m}$.
The DA computation order 3 is used.

\begin{lstlisting}[basicstyle={\footnotesize},tabsize=4,frame=shadowbox,mathescape=true]
INCLUDE 'COSY' ;

PROCEDURE RUN ;

VARIABLE Y1 4000 8 ; VARIABLE YT 4000 8 ;
VARIABLE XX0 4000 ; {Initial DA-valued x coordinate}
VARIABLE XX1 4000 ; {Final DA-valued x coordinate}
VARIABLE AA0 4000 ; {Initial DA-valued a coordinate}
VARIABLE AA1 4000 ; {Final DA-valued a coordinate}
VARIABLE NM 1 ; {DA variable size}
VARIABLE MU 1 ; {mu}
VARIABLE HPAR 4000 ; {h}

PROCEDURE FNC F X T ;
    {FUNCTION F(X,T) DEFINING THE ODEs OF MOTION
        X  COORDINATES VECTOR
        T  TIME (OR ARC LENGTH) }
    F(1) := X(2)/X(4) ; {r'}
    F(2) := -MU/X(1)/X(1)/X(4) + X(1)*X(4) ; {v_r'}
    F(3) := 1 ; {theta'}
    F(4) := -2*HPAR/X(1)/X(1)/X(1) * X(2)/X(4) ; {omega'}
ENDPROCEDURE ;

PROCEDURE RK4A N X0 X1 Y0 NS Y1 ;
    {FOURTH ORDER RUNGE KUTTA INTEGRATOR}
    VARIABLE I 1 ; VARIABLE J 1 ; VARIABLE T 1 ;
	VARIABLE Z1 NM 8 ; VARIABLE Z2 NM 8 ;
	VARIABLE Z3 NM 8 ; VARIABLE Z4 NM 8 ;
    VARIABLE Z5 NM 8 ; VARIABLE F NM 8 ;
    VARIABLE HS1 1 ;
    T := X0 ;
    HS1 := (X1-X0)/NS ;
    LOOP J 1 N ; Y1(J) := Y0(J) ; ENDLOOP ;
    LOOP I 1 NS ;
        FNC F Y1 T ;
        LOOP J 1 N ; Z1(J) := HS1*F(J) ; ENDLOOP ;
        LOOP J 1 N ; Z5(J) := Y1(J) + Z1(J)/2 ; ENDLOOP ;
        FNC F Z5 T+HS1/2 ;
        LOOP J 1 N ; Z2(J) := HS1*F(J) ; ENDLOOP ;
        LOOP J 1 N ; Z5(J) := Y1(J) + Z2(J)/2 ; ENDLOOP ;
        FNC F Z5 T+HS1/2 ;
        LOOP J 1 N ; Z3(J) := HS1*F(J) ; ENDLOOP ;
        LOOP J 1 N ; Z5(J) := Y1(J) + Z3(J) ; ENDLOOP ;
        FNC F Z5 T+HS1 ;
        LOOP J 1 N ; Z4(J) := HS1*F(J) ; ENDLOOP ;
        LOOP J 1 N ;
        	Y1(J) := Y1(J) + (1/6)*(Z1(J)+2*Z2(J)+2*Z3(J)+Z4(J)) ;
        ENDLOOP ;
        T := T + HS1 ;
    ENDLOOP ;
ENDPROCEDURE ;

PROCEDURE KEPLERODEPOLAR R00 PHI X0 A0 X1 A1 ;
    {This procedure calculates the transfer map of the electrostatic
    spherical deflector in (x,a) coordinates by integrating the
    ODEs of motion in laboratory coordinates using a Runge-Kutta
    integrator.
    Input parameters:
        R00    Radius of the circular reference orbit
        PHI    Central angle of tracked sector of the deflector
        X0     Initial x coordinate
        A0     Initial a coordinate
        X1     Final x coordinate
        A1     Final a coordinate }
    VARIABLE R0 NM 2 ; {Initial radius vector}
    VARIABLE W0 NM 2 ; {Initial velocity vector}
    VARIABLE R1 NM 2 ; {Final radius vector}
    VARIABLE W1 NM 2 ; {Final velocity vector}
    VARIABLE DOT NM ; {A scalar product}
    VARIABLE R0S NM ; {Initial radius magnitude}
    VARIABLE RS NM ; {Final radius magnitude}
    VARIABLE SIGMA0 NM ; {sigma_0}
    VARIABLE P NM ; {Focal parameter}
    VARIABLE F NM ; VARIABLE G NM ; {Lagrange coefficients F, G}
    VARIABLE FT NM ; VARIABLE GT NM ; {Lagrange coefficients FT, GT}
    VARIABLE V0 NM ; {Reference velocity}
    VARIABLE V1 NM ; {Initial velocity magnitude}
    VARIABLE V2 NM ; {Final velocity magnitude}
    VARIABLE CHIE 1 ; {Electric rigidity}
    PHI := PHI*DEGRAD ;
    V0 := CONS(SQRT(2*ETA)*CLIGHT) ;
    R0(1) := X0 + R00 ; R0(2) := 0 ;
    CHIE := AMU*RE(M0)*V0*V0 / Z0/EZERO ; {R00*CHIE}
    V1 := SQRT(V0^2-2*Z0*EZERO*CHIE*(1-R00/(R0(1)))/AMU/RE(M0)) ;
    W0(1) := A0*V0 ; W0(2) := SQRT(V1*V1-W0(1)*W0(1)) ; 
    MU := R00*CONS(V1)^2 ;
    HPAR := R0(1)*W0(2) ;
    YT(1) := R0(1) ;
    YT(2) := W0(1) ;
    YT(3) := 0 ;
    YT(4) := W0(2)/YT(1) ;
    {RK8A 4 0 PHI YT HS Y1 1e-5 RESCODE ;}
    RK4A 4 0 PHI YT 1e6 Y1;
    X1 := Y1(1) - R00 ;
    V2 := SQRT(Y1(2)*Y1(2)+Y1(1)*Y1(1)*Y1(4)*Y1(4)) ;
    A1 := Y1(2) / CONS(V2) ;
ENDPROCEDURE ;

OV 3 1 0 ; 
RP 1 1 1 ; 

NM := 4000 ;
XX0 := DA(1) ; AA0 := DA(2) ;

KEPLERODEPOLAR 1 45 XX0 AA0 XX1 AA1 ;
WRITE 6 'TRANSFER MAP OBTAINED IN LAB COORDINATES' ;
WRITE 6 'BY INTEGRATION OF THE ODEs' ;
WRITE 6 'X_f' XX1 'A_f' AA1 ;

ENDPROCEDURE ; RUN ; END ; 
\end{lstlisting}

To integrate the ODEs of motion in polar laboratory coordinates to
calculate the transfer map of a $45\textdegree$ sector of the electrostatic
cylindrical deflector in $\left(x,a\right),$ the code has to be adjusted
according to the changes in equations as described in Sub-Sec. \ref{sec:ECL}.
The change in eq. \ref{eq:ODEs-2CL} is to be reflected in the $\mathtt{F(2)}$
line in $\mathtt{PROCEDURE}$ $\mathtt{FNC}$ as listed below. 

\begin{lstlisting}[basicstyle={\footnotesize},tabsize=4,frame=shadowbox,mathescape=true]
PROCEDURE FNC F X T ;
    {FUNCTION F(X,T) DEFINING THE ODEs OF MOTION
        X  COORDINATES VECTOR
        T  TIME (OR ARC LENGTH) }
    F(1) := X(2)/X(4) ; {r'}
    F(2) := -MU/X(1)/X(4) + X(1)*X(4) ; {v_r'}
    F(3) := 1 ; {theta'}
    F(4) := -2*HPAR/X(1)/X(1)/X(1) * X(2)/X(4) ; {omega'}
ENDPROCEDURE ;
\end{lstlisting}
The change in $v_{i}$ in eq. \ref{eq:v1-velocityCL} is to be reflected
in the $\mathtt{V1}$ line in $\mathtt{PROCEDURE}$ $\mathtt{KEPLERODEPOLAR}$
as listed below. The rest of the code can stay the same.

\begin{lstlisting}[basicstyle={\footnotesize},tabsize=4,frame=shadowbox,mathescape=true]
PROCEDURE KEPLERODEPOLAR R00 PHI X0 A0 X1 A1 ;
    {This procedure calculates the transfer map of the electrostatic
    cylindrical deflector in (x,a) coordinates by integrating the
    ODEs of motion in laboratory coordinates using a Runge-Kutta
    integrator.
    ...(continue)...
    CHIE := ...(the same as above)...
    V1 := SQRT(V0^2-2*Z0*EZERO*CHIE*R00*LOG(R0(1)/R00)/AMU/RE(M0)) ;
    W0(1) := ...(the same as above)...
    ...(continue)...
ENDPROCEDURE ;
\end{lstlisting}
\pagebreak{}

\section{\emph{Mathematica} Code for Integration of the ODEs in Laboratory
Coordinates\label{sec:code-ODE-Mathematica}}

The following is a \emph{Mathematica} notebook that integrates the
ODEs of motion of a particle with specified initial beamline coordinates
$\left(x,a\right)$ through the electrostatic spherical deflector
in polar laboratory coordinates, plots one turn of the orbit, and
outputs the final beamline coordinates at true anomaly difference
$\theta=45\textdegree$. The inner sphere of the electrostatic spherical
deflector is charged to result in a circular reference orbit of radius
$r_{0}=1\:\mathrm{m}$.

\begin{lstlisting}[basicstyle={\footnotesize},tabsize=4,frame=shadowbox,mathescape=true]
(* Integration of the ODEs of motion for a charged \ 
particle of kinetic energy 1 MeV, mass 1 amu, and \
charge 1 e in an electrostatic spherical deflector \
with the inner sphere charge such that the particle \
would have a circular reference orbit of radius 1 m. *)
v00 = 13891388.79714028; (* Reference velocity *)
r00 = 1; (* Reference orbit radius *)
x = -0.5; (* Initial x beamline coordinate *)
a = 0.5; (* Initial a beamline coordinate *)
r0 = {x + r00, 0, 0}; (* Initial radius vector *)
v0 = {a v00, v00 Sqrt[1 - a^2], 0}; (* Initial velocity vector *)
mu = r00  v00^2; (* mu *)
vr0 = v0[[1]]; 
rr0 = r0[[1]];
omega0 = v0[[2]]/rr0; (* omega_0 *)
theta0 = 0; (* Initial polar angle *)
system = { (* ODEs and the initial condition *)
   vr'[t] == -mu/(omega[t] (rr[t])^2) + rr[t] omega[t],
   rr'[t] == vr[t]/omega[t],
   omega'[t] == -2 rr0 v0[[2]] vr[t]/(omega[t] (rr[t])^3),
   theta'[t] == 1,
   vr[0] == vr0,
   rr[0] == rr0,
   omega[0] == omega0,
   theta[0] == theta0
   };
sol = NDSolve[system, {vr, rr, omega, theta}, {t, 0, 2 \[Pi]}] ;
rsol[t_] = First[Evaluate[rr[t] /. sol]];
 (* Radius magnitude as a function of polar angle theta *)
vsol[t_] = First[Evaluate[vr[t] /. sol]];
 (* Velocity magnitude as a function of polar angle theta *)
Print[PolarPlot[
   rsol[t], {t, 0, 2 \[Pi]}]]; (* Plot one turn of the orbit *)
x1[t_] = rsol[t] - r00; (* x coordinate at polar angle theta *)
a1[t_] = vsol[t]/v00; (* a coordinate at polar angle theta *)
Print["x_f=", x1[Pi/4]]; 
Print["a_f=", a1[Pi/4]];
\end{lstlisting}
\pagebreak{}

\section{\emph{COSY INFINITY} Code for Transfer Map Calculation Using Lagrange
Coefficients in Laboratory Coordinates\label{sec:code-Kepler-COSY}}

The following is a \emph{COSY INFINITY} code that calculates the transfer
map of a $45\textdegree$ sector of the electrostatic spherical deflector
in $\left(x,a\right)$ beamline coordinates using Lagrange coefficients
in polar laboratory coordinates as described in Sec. \ref{sec:Kepler}.
The DA computation order 3 is used.

\begin{lstlisting}[basicstyle={\footnotesize},tabsize=4,frame=shadowbox,mathescape=true]
INCLUDE 'COSY' ;

PROCEDURE RUN ;

VARIABLE XX0 4000 ; {Initial DA-valued x coordinate}
VARIABLE XX1 4000 ; {Final DA-valued x coordinate}
VARIABLE AA0 4000 ; {Initial DA-valued a coordinate}
VARIABLE AA1 4000 ; {Final DA-valued a coordinate}
VARIABLE NM 1 ; {DA variable size}

PROCEDURE KEPLERANALYTICPOLAR R00 PHI X0 A0 X1 A1 ;
   VARIABLE R0 NM 2 ; {Initial radius vector}
   VARIABLE W0 NM 2 ; {Initial velocity vector}
   VARIABLE R1 NM 2 ; {Final radius vector}
   VARIABLE W1 NM 2 ; {Final velocity vector}
   VARIABLE MU NM ; {mu}
   VARIABLE DOT NM ; {A scalar product}
   VARIABLE R0S NM ; {Initial radius}
   VARIABLE RS NM ; {Final radius}
   VARIABLE SIGMA0 NM ; {sigma_0}
   VARIABLE P NM ; {Focal parameter p}
   VARIABLE F NM ; VARIABLE G NM ; {Lagrange coefficients F and G}
   VARIABLE FT NM ;
   VARIABLE GT NM ; {Lagrange coefficients F_t and G_t}
   VARIABLE V0 NM ; {Initial velocity at zero potential}
   VARIABLE V1 NM ; {Initial velocity}
   VARIABLE V2 NM ; {Final velocity}
   VARIABLE CHIE 1 ; {Electric rigidity}
   PHI := PHI*DEGRAD ;
   V0 := CONS(SQRT(2*ETA)*CLIGHT) ;
   R0(1) := X0 + R00 ; R0(2) := 0 ;
   CHIE := AMU*RE(M0)*V0*V0 / Z0/EZERO ; {R00*CHIE}
   V1 := SQRT(V0^2 - 2*Z0*EZERO*CHIE*(1-R00/(R0(1)))/AMU/RE(M0)) ;
   W0(1) := A0*V0 ;
   W0(2) := SQRT(V1*V1-W0(1)*W0(1)) ; 
   MU := R00*CONS(V1)^2 ;
   DOT := R0(1)*W0(1) + R0(2)*W0(2) ;
   SIGMA0 := DOT/SQRT(MU) ;
   P := ((R0(1)*R0(1)+R0(2)*R0(2))*(W0(1)*W0(1)+W0(2)*W0(2))
         -DOT*DOT) / MU ;
   R0S := R0(1) ;
   RS := P*R0S / (R0S+(P-R0S)*COS(PHI)-SQRT(P)*SIGMA0*SIN(PHI)) ;
   F := 1 - (RS/P)*(1-COS(PHI)) ;
   G := (RS*R0S/SQRT(MU*P)) * SIN(PHI) ;
   FT := (SQRT(MU)/R0S/P) * (SIGMA0*(1-COS(PHI))-SQRT(P)*SIN(PHI)) ;
   GT := 1 - (R0S/P)*(1-COS(PHI)) ;
   R1(1) := F*R0(1) + G*W0(1) ; R1(2) := F*R0(2) + G*W0(2) ; 
   W1(1) := FT*R0(1) + GT*W0(1) ; W1(2) := FT*R0(2) + GT*W0(2) ; 
   X1 := (COS(PHI)*R1(1)+SIN(PHI)*R1(2)) - R00 ;
   V2 := SQRT(W1(1)*W1(1)+W1(2)*W1(2)) ;
   A1 := (COS(PHI)*W1(1)+SIN(PHI)*W1(2)) / CONS(V2) ;
ENDPROCEDURE ;

OV 3 1 0 ; 
RP 1 1 1 ; 

NM := 4000 ;
XX0 := DA(1) ; AA0 := DA(2) ;

KEPLERANALYTICPOLAR 1 45 XX0 AA0 XX1 AA1 ;
WRITE 6 'TRANSFER MAP OBTAINED IN LAB COORDINATES' ;
WRITE 6 'USING LAGRANGE COEFFICIENTS' ;
WRITE 6 'X_f' XX1 'A_f' AA1 ;

ENDPROCEDURE ; RUN ; END ;
\end{lstlisting}
\pagebreak{}

\section{\emph{COSY INFINITY} Codes for Transfer Map Calculation of the Built-in
Electrostatic Deflector Elements\label{sec:code-ESP}}

The following is a \emph{COSY INFINITY} code to compute the transfer
map of a $45\textdegree$ sector of the electrostatic spherical deflector.
The voltages on the plates of the electrostatic deflectors are picked
to result in a circular reference orbit of radius $r_{0}=1\:\mathrm{m}$.
The DA computation order 3 is used.

\begin{lstlisting}[basicstyle={\footnotesize},tabsize=4,frame=shadowbox,mathescape=true]
INCLUDE 'COSY' ;

PROCEDURE RUN ;

OV 3 1 0 ; 
RP 1e-7 1 1 ;

UM ;
ESP 1 45 0.1 ; {Electrostatic spherical deflector}

WRITE 6 'TRANSFER MAP OF COSY INFINITY''S ESP ELEMENT' ;
WRITE 6 'X_f' MAP(1) 'A_f' MAP(2) ;

ENDPROCEDURE ; RUN ; END ; 
\end{lstlisting}

To compute the transfer map of a $45\textdegree$ sector of the electrostatic
cylindrical deflector, the $\mathtt{ESP}$ call line above is to be
replaced by a $\mathtt{ECL}$ call.
\begin{lstlisting}[basicstyle={\footnotesize},tabsize=4,frame=shadowbox,mathescape=true]
...(the same as above)...
ECL 1 45 0.1 ; {Electrostatic cylindrical deflector}
...(the same as above)...
\end{lstlisting}
\pagebreak{}

\section{\emph{GIOS} Input for Transfer Map Calculation of the Electrostatic
Deflector Elements\label{sec:code-GIOS}}

The following is a \emph{GIOS} input to compute the transfer map of
a $45\textdegree$ sector of the electrostatic spherical deflector.
The computation order 2 is used. We note that the $\mathtt{SC}$ command
is not essential for our purpose to compare with the transfer maps
computed by \emph{COSY INFINITY}. 

\begin{lstlisting}[basicstyle={\footnotesize},tabsize=4,frame=shadowbox,mathescape=true]
R P 1 1 1 ; Reference Particle
C O 2 2 ;   Calculation Order
S C ;       Symplectic Coordinate system for map printout
;
E S 1 45 0.01 1 -1 1 ; Electrostatic Sector, spherical
;
END ;
END ;
\end{lstlisting}

To compute the transfer map of a $45\textdegree$ sector of the electrostatic
cylindrical deflector, the description of the $\mathtt{ES}$ command
is modified, and a \emph{GIOS} input is as follows. We note again
that the $\mathtt{SC}$ command is not essential for our purpose.

\begin{lstlisting}[basicstyle={\footnotesize},tabsize=4,frame=shadowbox,mathescape=true]
R P 1 1 1 ; Reference Particle
C O 2 2 ;   Calculation Order
S C ;       Symplectic Coordinate system for map printout
;
E S 1 45 0.01 0 0 0 ; Electrostatic Sector, cylindrical
;
END ;
END ;
\end{lstlisting}


\begin{thebibliography}{1}

\bibitem{battin}
Richard~H. Battin.
\newblock {\em An Introduction to the Mathematics and Methods of Astrodynamics,
  Revised Edition}.
\newblock American Institute of Aeronautics and Astronautics, Reston, VA, rev.
  edition, 1999.

\bibitem{Berz1999}
Martin Berz.
\newblock {\em Modern Map Methods in Particle Beam Physics}.
\newblock Advances in Imaging and Electron Physics. Academic Press, San Diego,
  CA, 1999.

\bibitem{COSY-BPM-10}
Martin Berz and Kyoko Makino.
\newblock {COSY INFINITY 10.0 Beam Physics Manual}.
\newblock {MSU Report MSUHEP 151103-rev}, Department of Physics and Astronomy,
  Michigan State University, East Lansing, MI 48824, 2017.
\newblock See also http://cosyinfinity.org.

\bibitem{COSY}
Kyoko Makino and Martin Berz.
\newblock {COSY INFINITY Version 9}.
\newblock {\em Nucl. Instr. Meth. Phys. Res. A}, 558(1):346--350, 3 2006.

\bibitem{GIOS}
H.~Wollnik, J.~Brezina, and M.~Berz.
\newblock {GIOS-BEAMTRACE -- A Program Package to Determine Optical Properties
  of Intense Ion Beams}.
\newblock {\em Nucl. Instr. Meth. Phys. Res. A}, 258(3):408--411, 1987.

\bibitem{wollnik-nimsys}
Hermann Wollnik and Martin Berz.
\newblock {Relations Between Elements of Transfer Matrices Due to the Condition
  of Symplecticity}.
\newblock {\em Nucl. Instr. Meth. Phys. Res. A}, 238(1):127--140, 1985.

\end{thebibliography}
\end{document}